\documentstyle[twoside,fleqn,espcrc2,psfig]{article}

\newcommand{\gsim}{\lower.7ex\hbox{$
    \;\stackrel{\textstyle>}{\sim}\;$}}
\newcommand{\lsim}{\lower.7ex\hbox{$
    \;\stackrel{\textstyle<}{\sim}\;$}}
\def\Li2#1{{\rm Li}_2\left(#1\right)}
\newcommand{\be}{\begin{eqnarray}}
\newcommand{\ee}{\end{eqnarray}}
\def\dd{{\rm d}}

\title{Lepton anomalous magnetic moments -- a theory update}

\author{Andrzej Czarnecki and William J. Marciano\address{Physics
Department\\ 
Brookhaven National Laboratory\\Upton, New York 11973}%
        \thanks{This research has been supported by the
U.S.~Department of Energy grant DE-AC02-98CH10886.}
}
       
\begin{document}
 
\begin{abstract}
Standard Model contributions to the electron, muon, and tau lepton
anomalous magnetic moments, $a_l=(g_l-2)/2$, are reviewed and updated.
The fine structure constant is obtained from the electron $g_e-2$ and
used to refine the QED contribution to the muon $g_\mu-2$.  Recent
advances in electroweak and hadronic effects on $g_\mu-2$ are
summarized.  Examples of ``New Physics'' probed by the $a_\mu$
Brookhaven experiment E821 are outlined. The prediction for $a_\tau$
is also given.
\end{abstract}

\maketitle

\section{Introduction and summary}

\vspace*{-100mm}
\begin{minipage}{156mm}
\begin{flushright}
BNL-HET-98/43\\
hep-ph/9810512\\
October 1998
\end{flushright}
\end{minipage}
\vspace*{89mm}

During the past few years, there has been growing interest in leptonic
anomalous magnetic moments.  On the experimental side, the
extraordinary measurements of $a_e \equiv (g_e-2)/2$ at the University
of Washington currently provide the best determination of the fine
structure constant, $\alpha$, when compared with theoretical
predictions:
\be
\alpha^{-1} = 137.035\;999\;59(38)(13).
\ee
  Similarly, a new effort in progress at Brookhaven
National Lab (Experiment E821) aims to improve the measurement of
$a_\mu$ by a factor of 20 or better. Although not competitive with
$a_e$ in precision, $a_\mu$ is much more sensitive to electroweak loop
effects as well as ``New Physics'' which give contributions $\sim
m_l^2$, i.e.~4$\times 10^4$ enhancement in $a_\mu$ relative to $a_e$.
Besides being able to observe the electroweak loop corrections to
$a_\mu$ predicted by the Standard Model, E821 is capable of detecting
the presence of ``New Physics.''  For example, supersymmetry loop
effects can potentially provide a large contribution to $a_\mu$. If a
significant deviation from Standard Model expectations is observed,
supersymmetry is likely to provide the leading candidate explanation. 

To exploit experimental progress requires detailed calculations of the
Standard Model contributions to $a_e$ and $a_\mu$.  QED computations
at the 4 and 5 loop level, hadronic vacuum polarization and other loop
effects, 1 and 2 loop electroweak effects, and ``New Physics''
contributions must be throughly scrutinized and refined. That effort
challenges our theoretical tools and abilities. It provides an
important synergism between theory and experiment. 

\begin{table}[htb]
\setlength{\tabcolsep}{1.0pc}
\caption{Standard Model  predictions and experimental results for the
 leptonic anomalous magnetic moments.}
\label{tab1}
\begin{tabular*}{0.47\textwidth}{ll}  \hline  
$a_{e}^{\rm theory}=0.001\,  159\,  652 \, 153\, 5(240)$    & \\ 
$a_{e^-}^{\rm exp\phantom{wy}} =0.001\, 159\, 652 \,   188\, 4(43)$ &
\cite{Dehmelt87} \\  
$a_{e^+}^{\rm exp\phantom{wy}} =0.001\, 159\, 652 \,   187\, 9(43)$ &
\cite{Dehmelt87} \\  
\hline
$a_{\mu}^{\rm theory}=0.001\, 165\, 915 \, 96(67)$    & \\ 
$a_{\mu}^{\rm exp\phantom{wy}}   =0.001\, 165\, 923 \, 50(730)$ & 
                                            \cite{PDG98,VernonBulg}\\ 
\hline
$a_{\tau}^{\rm theory}=0.001\, 176\, 9(4)$    & \\ 
$a_{\tau}^{\rm exp\phantom{wy}}   =0.004(35)$ &  \cite{Taylor98} \\ 
\hline
\end{tabular*}
\end{table}

In this paper we update some of the recent progress in theoretical
calculations of $a_e$ and $a_\mu$ and thereby provide a status report
(see Table \ref{tab1}). In the case of $a_\mu$, we briefly describe a
few examples of the ``New Physics'' sensitivity of E821 underway at
Brookhaven. Comparison
of the present experimental results with the Standard Model prediction
gives the following 95\% C.L. bound on ``New Physics'' contributions
to the muon anomalous magnetic moment
\be
-710 \times 10^{-11} < a_\mu^{\rm New \; Physics} < 2210 \times
10^{-11}.
\ee
For completeness, we also give an updated prediction for
$a_\tau$, even though an experimental measurement of that quantity is
far from current capabilities.

\section{Electron}
To match the present experimental precision one has to include the
following QED contributions to $a_e=(g_e-2)/2$
\be
\lefteqn{a_e^{\rm QED} 
 = \sum_{n=1}^4 A_n \left( {\alpha\over \pi}\right)^n}
\nonumber \\ &&\hspace*{-7mm}
 + \left[B_2(e,\mu)+B_2(e,\tau)\right] \left({\alpha\over\pi}\right)^2
 + B_3(e,\mu) \left({\alpha\over\pi}\right)^3
\ee
where $B_n(l,l')$  describe the contributions of  loops containing
lepton $l'$ to $a_l$, while $A_n$ contain pure QED contributions
\cite{Schwinger48,som57,pet57a,Laporta:1996mq,Kinoshita98}: 
\be
A_1 &=& {1\over 2}
\nonumber \\
A_2 &=& {3\over 4}\zeta_3 -{\pi^2\over 2} \ln 2+{\pi^2\over 12}
+{197\over 144}
\nonumber \\
&\approx & -0.3284789656
\nonumber \\
A_3 &=& {83\over 72} \pi^2 \zeta_3-{215\over 24} \zeta_5
 -{239\over 2160} \pi^4+{139\over 18} \zeta_3
\nonumber \\ &&
 +{25\over 18} 
\left[24{\rm Li}_4\left({1\over 2}\right) +\ln^4 2-\pi^2 \ln^2
                         2\right]
\nonumber \\ &&
 -{298\over 9} \pi^2 \ln 2
 +{17101\over 810} \pi^2+{28259\over 5184}
\nonumber \\ & \approx & 1.1812415
\nonumber \\
A_4 &=& -1.5098(384) 
\label{eq:eqed}
\ee
and
\be
B_2(e,\mu) &=& 5.197 \times 10^{-7}
\nonumber \\
B_2(e,\tau) &=& 1.838 \times 10^{-9}
\nonumber \\
B_3(e,\mu) &\simeq & -7.3739 \times 10^{-6}.
\ee
$B_2(e,l)$ describe loops with the lepton $l$ inserted in the Schwinger
diagram; they are calculated using
\begin{eqnarray}
B_2(e,l) &=&{1\over 3} \int_{4m_l^2}^\infty \dd s
 \sqrt{s-4m_l^2\over s}{s+2m_l^2\over s^2}
\nonumber \\ && \times
 \int_0^1 \dd x {x^2(1-x)\over x^2+(1-x){s\over m_e^2}}
\end{eqnarray}
$B_3(e,\mu)$  is a sum of three groups of diagrams
\cite{Bar75,Laporta:1993ju,Laporta:1993pa}:
\be 
B_3(e,\mu)&=&A^{(4,2)}(m_\mu/m_e)
+B^{(2,4)}(m_\mu/m_e)
\nonumber \\ &&
+B_3^{\gamma\gamma}(e,\mu) . 
\ee
They describe, respectively, diagrams with either two muon loops or photon
corrections within a single muon loop; and 
the remaining diagrams with a single muon loop (they contain either
a photon not attached to the muon loop or and electron loop). 
Finally, $B_3^{\gamma\gamma}(e,\mu)$ is a contribution of
light-by-light scattering with a muon loop \cite{Laporta:1993pa}.
Their numerical values are 
\be
\lefteqn{
A^{(4,2)}(m_\mu/m_e)+B^{(2,4)}(m_\mu/m_e) \approx 
}
\nonumber\\
&&
\left({m_e\over m_\mu}\right)^2
\left[-{23\over 135}\ln\left({m_\mu\over m_e}\right)-{2\over 45}\pi^2
+{10117\over 24300}\right]
\nonumber\\
&&
+\left({m_e\over m_\mu}\right)^4
\left[
{19\over 2520} \ln^2\left({m_\mu\over m_e}\right)
\right.
\nonumber\\
&&
-{14233\over 132300}\ln\left({m_\mu\over m_e}\right)
+{49\over 768}\zeta_3 -{11\over 945}\pi^2
\nonumber\\
&&
\left.
+{2976691\over 296352000}\right] 
\approx -0.000021768
\ee
and 
\be
\lefteqn{B_3^{\gamma\gamma}(e,\mu) \approx 
\left({m_e\over m_\mu}\right)^2
\left[{3\over 2}\zeta_3-{19\over 16}\right]
}
\nonumber\\
&&
+\left({m_e\over m_\mu}\right)^4
\left[
-{161\over 810} \ln^2\left({m_\mu\over m_e}\right)
\right.
\nonumber \\
&& 
-{16189\over 48600}\ln\left({m_\mu\over m_e}\right)
+{13\over 18}\zeta_3 -{161\over 9720}\pi^2
\nonumber \\
&& \left. 
-{831931\over 972000}\right]
 \approx 0.0000143945
\ee

The hadronic contributions arise from vacuum polarization insertion in
the Schwinger diagram \cite{Davier:1998si}, in the two-loop QED
diagrams \cite{Krause:1997rf}, and from the hadronic light-by-light
diagram, estimated as $a_\mu^{\rm had}(\mbox{light-by-light})
m_e^2/m_\mu^2$; the result is
\be
a_e^{\rm had} = 1.63(3)\times 10^{-12}, \qquad 
\ee
The electroweak contribution up to two loops is \cite{CKM96}
\be
a_e^{\rm EW} =  0.030\times 10^{-12}.
\ee

In total, the current Standard Model prediction for $a_e$ is given by 
\be
\lefteqn{a_e = 0.5{\alpha\over \pi} -0.328478444
\left({\alpha\over \pi}\right)^2}
\nonumber
\\
&&\hspace*{-5mm} +1.181234\left({\alpha\over \pi}\right)^3
-1.5098\left({\alpha\over \pi}\right)^4+1.66\times 10^{-12}
\nonumber\\
\ee
where the last term consists of the electroweak and hadronic
contributions, whose dependence on $\alpha$ can be neglected for the
purpose of $\alpha$ determination at the present level of accuracy.  
The QED part of the expansion appears to converge very well, with
alternating signs and no considerable growth of the coefficients. 
We can use the average of the latest experimental results for $a_e$
\cite{Dehmelt87}, 
\be
a_{e^-}^{\rm exp} &=& 1159652188.4(4.3)\times 10^{-12},
\nonumber \\ 
a_{e^+}^{\rm exp} &=& 1159652187.9(4.3)\times 10^{-12},
\ee
to deduce a value of $\alpha$:
\be
\alpha^{-1} = 137.035\;999\;59(38)(13)
\ee
where the first error comes from the uncertainty $\Delta a^{\rm
exp}_e=4.3\times 10^{-12}$ (dominated by systematic cavity effects),
and the second from the theoretical uncertainty in $A_4$.

The next most precise method of determining $\alpha$ is based on the
Quantum Hall Effect (for a review see \cite{Kinoshita:1996vz} and
references therein), where one finds  \cite{Jeffery}
\be
\alpha^{-1}(qH) = 137.036\;003\;70(270).
\label{eq:qH}
\ee
We have used that value of $\alpha$ in obtaining the prediction for
$a_e^{\rm theory}$ in Table \ref{tab1}.

Although there is currently a small discrepancy (1.5$\sigma$) between
$\alpha^{-1}(a_e)$ and $\alpha^{-1}(qH)$, the overall level of
agreement is truly impressive and represents a triumph for
QED. However, the good agreement is not a particularly severe
constraint on ``New Physics.''  If one assumes that such effects
contribute $\Delta a_e^{\rm New\; Physics}\sim m_e^2/\Lambda^2$, where
$\Lambda$ is the scale of ``New Physics,'' then a comparison of
$\alpha^{-1}(a_e)$ and $\alpha^{-1}(qH)$ is currently sensitive to
$\Lambda\lsim 100$ GeV.  To probe the much more interesting
$\Lambda\sim {\cal O}$(TeV) region would require an order of magnitude
improvement in $a_e$ (possibly feasible; see \cite{Gab94} for a
discussion), an improved calculation of $A_4$, and a better
independent direct measurement of $\alpha^{-1}$.  The last requirement
could be met by improving $\alpha^{-1}(qH)$ by 2 orders of magnitude
or perhaps more likely combining the already precisely measured
Rydberg constant with a better $m_e$ determination.

\section{Muon}
\subsection{QED contribution}
Because of the presence of virtual electron loops, higher
order QED contributions 
to $a_\mu$ are enhanced in comparison with $a_e$.  At present we need
5 terms of the expansion in $\alpha$:
\be
a_\mu^{\rm QED} = \sum_{n=1}^5 C_n\left( {\alpha\over \pi}\right)^n
\ee
with
\be
C_1 &=& A_1 = 0.5,\nonumber \\
C_2 &=& A_2 + a_1(m_e/m_\mu) + a_2(m_\tau/m_\mu) 
\nonumber \\
&=&
0.765\;857\;388(44),
\nonumber \\
C_3 &=& A_3+C_3^{\gamma\gamma}(e)+C_3^{\gamma\gamma}(\tau)
 +C_3^{\rm vac.\ pol.}(e) 
\nonumber \\ &&
+C_3^{\rm vac.\ pol.}(\tau)
 +C_3^{\rm vac.\ pol.}(e,\tau) 
\nonumber \\
&=& 24.050\;509(2),
\nonumber \\
C_4 &=& A_4+ 127.55(41) = 126.04(41),
\nonumber \\
C_5&=&  930(170).
\ee
where $a_{1,2}$ are given in \cite{Li:1993xf,Elend66}.
Taking $m_\mu/m_e=206.768273(24)$ \cite{Liu98} and
$m_\tau=1777.05(26)$ MeV, $m_e=0.51099907(15)$ MeV  \cite{PDG98}
we find
\be
a_1(m_e/m_\mu)&=&   1.094258294(37), 
\nonumber\\
a_2(m_\tau/m_\mu)&=& 0.000078059(23)
\ee
For the evaluation of $C_2$ the errors in $a_{1,2}$ have been added in
quadrature. 

In $C_3$ we have contributions from light-by-light scattering diagrams
with $e$ and $\tau$ loops, and vacuum polarization diagrams with either $e$,
or $\tau$, or both types of loops.  For light-by-light we use the
formulas of \cite{Laporta:1993pa}.  For vacuum polarization we use
\cite{Laporta:1993ju}, with exception of the mixed $e-\tau$ diagram
which we evaluate numerically using the kernel from
\cite{Krause:1997rf}.  With updated values of $m_\mu/m_e$ and $m_\tau$
we find
\be
C_3^{\gamma\gamma}(e)&=& \phantom{-} 20.9479246(7)
 \nonumber \\
C_3^{\gamma\gamma}(\tau) &=&\phantom{-2} 0.0021428(7)
 \nonumber \\
C_3^{\rm vac.\ pol.}(e) &=& \phantom{-2}1.9204551(2)
 \nonumber \\
C_3^{\rm vac.\ pol.}(\tau) &=&\;\,            -0.0017822(4)
 \nonumber \\
C_3^{\rm vac.\ pol.}(e,\tau) &=&\phantom{-2}0.0005276(2)
\ee
Even adding all errors in $C_3$ linearly, we end up with the
uncertainty 20 times smaller than the previous update \cite{CKM96}.
This is because of the availability of the analytical result for $A_3$
\cite{Laporta:1996mq}, whose error previously dominated.

For $C_4$ we use the difference between the muon and electron
coefficients found in \cite{Kinoshita:1993pq}, and the latest $A_4$
value; with errors added in quadrature we get
\be
C_4 = 127.55(41)+A_4 = 126.04(41).
\ee

The first partial evaluation of $C_5$ was performed in
\cite{Kinoshita:1990wp}, where one-loop vacuum polarization insertions
in the lowest order light-by-light diagram were computed, giving a
contribution of about 570 to $C_5$.  Those are the diagrams with the
maximal power of $\ln (m_\mu/m_e)$.  However, these are not the only
strongly enhanced diagrams.  As had been shown in
\cite{Elkhovskii:1989cn} the light-by-light diagrams, in which the
electron loop is connected with $2n+1$ photons to the muon loop,
contains $\pi^{2n}\ln(m_\mu/m_e)$.  The numerical coefficient of such
term in $C_5$ has been calculated in \cite{Mil89}.  Some other
diagrams, including two-loop vacuum polarization insertions, were
estimated in \cite{Karshenboim:1993rt}; we adopt the value of $C_5$ 
from this paper.

Some diagrams contributing to $C_5$ have been evaluated analytically
in \cite{Kataev:1992cp,Laporta:1994md}.  Other estimates of $C_5$ have
also been made, using various methods including renormalization group
\cite{Kataev:1992cp,Kataev:1995rw} and Pad\'e approximation 
\cite{Ellis:1994qf}.  A class of contributions has even been computed
to all orders \cite{Broadhurst:1993si}.

We now understand the ratio of the growth of the $C_i$ coefficients
describing the QED contribution to the muon $g-2$.  In the terms
calculated so far, the characteristic increase of the coefficients is
of the order of 10 for each power of $\alpha/\pi$.  The primary reason
of the growth is an extra factor of $\ln(m_\mu/m_e) \simeq 5.3$ due to
electron loops inserted in the photon propagators, further enhanced by
combinatorial factors (this also explains why the higher order terms
have the same sign as the first one; the perturbative series tries to
compensate for the fact that we are using a too small scale, and
therefore too small value of $\alpha$).  Another source of increase is
due to the factor of $\pi^2 \simeq 9.9$, which accompanies each two
additional photon rungs added to the light-by-light diagrams.  This
reasoning justifies the truncation of the perturbative series at the
fifth term.  Our final estimate of the QED contribution to $a_\mu$ is
\be
a_\mu^{\rm QED} = 116\;584\;705.6(2.9) \times 10^{-11}.
\ee
The error has been estimated by linearly adding roughly equal
contributions from the uncertainty in $\alpha$ and $C_{4,5}$, and a
small number estimating the higher order terms in the QED series for
$a_\mu^{\rm QED}$.

\subsection{Hadronic contributions}
The bulk of hadronic contributions comes from vacuum polarization
insertion in the Schwinger diagram and can be calculated using
experimental data on $e^+e^-\to$ hadrons or hadronic $\tau$ decays,  
and dispersion relations (see \cite{GRaf} and \cite{Jeg95} where further
references can be found).  The most recent evaluation
\cite{Davier:1998si} gives
\be
a_\mu^{\rm had}(\mbox{4th order}) = 6924(62) \times 10^{-11}
\ee
That study relies in part on theoretical assumptions which are 
subject to some debate. For example, a preliminary update of the
analysis \cite{Jeg95} arrives at a larger error estimate of $119\times
10^{-11}$ \cite{Simon}.
 
In addition, one has to include the vacuum polarization insertion in
all two-loop QED diagrams \cite{Alemany:1997tn,Krause:1997rf},
\be
a_\mu^{\rm had}(\mbox{6th order}) = -100(6)\times 10^{-11},
\ee
and hadronic light-by-light diagram \cite{light}
for which we take an average \cite{Alemany:1997tn} of
the values given in \cite{Hayakawa:1998rq,Bijnens:1996xf}
\be
a_\mu^{\rm had}(\mbox{light-by-light}) = -85(25)\times 10^{-11}.
\ee
The total hadronic contribution is
\be
a_\mu^{\rm had} = 6739(67)\times 10^{-11}.
\ee
Its error dominates the present theoretical prediction of $a_\mu$.

It would be very valuable to have at least an independent estimate of
the hadronic light-by-light contributions from lattice QCD.  Such a
study might be undertaken in the near future \cite{OhtaPC}.  Also,
ongoing studies of $e^+e^-\to$ hadrons and hadronic tau decays could
further reduce the theoretical uncertainty. A goal of $\pm 40\times
10^{-11}$ or smaller is well matched to the prospectus of experiment
E821 at Brookhaven which aims for that level of experimental accuracy.

\subsection{Electroweak contributions}
At the one loop level, the Standard Model predicts 
\cite{fls72,Jackiw72,ACM72,Bars72,Bardeen72}
\begin{eqnarray}
\lefteqn{a_\mu^{\rm EW}(\rm 1\,loop) =
{5\over 3}{G_\mu m_\mu^2\over 8\sqrt{2}\pi^2}}
\nonumber\\ && \times
\left[1+{1\over 5}(1-4s_W^2)^2
+ {\cal O}\left({m_\mu^2 \over M^2}\right) \right]
\nonumber \\
&& \approx 195 \times 10^{-11}
\label{eq:oneloop}
\end{eqnarray}
where $G_\mu = 1.16639(1) \times 10^{-5}$ GeV$^{-2}$, $M=M_W$ or
$M_{\rm Higgs}$, and the weak mixing angle
$\sin^2\theta_W\equiv s_W^2 = 1-M_W^2/M_Z^2=0.224$.  We can safely
neglect the ${\cal O}\left({m_\mu^2 / M^2}\right)$ terms in
(\ref{eq:oneloop}).

Two-loop corrections \cite{CKM95,CKM96,KKSS,Peris:1995bb}
\be
a_\mu^{\rm EW} (\mbox{2-loop}) = -44(4)\times 10^{-11},
\ee
decrease the
electroweak contribution by about 23\%, bringing it to 
\be
a_\mu^{\rm EW} = 151(4)\times 10^{-11}.
\ee
This decrease is mainly due to the large logarithmic terms $\sim \ln
(M_Z^2/m_\mu^2)$. 
These leading logs have recently been resummed to
all orders in $\alpha$ \cite{Degrassi:1998es}.  They found that the
leading logs in order ${\cal O}(\alpha^3)$ increase the EW effects
by a very small amount, $a_\mu^{\rm EW}(\mbox{3-loop, LL}) = 0.5\times
10^{-11}$ which is within the error.

The complete Standard Model prediction for $a_\mu$ is 
\be
a_\mu^{\rm SM} &=& a_\mu^{\rm QED} + a_\mu^{\rm had} + a_\mu^{\rm EW}
\nonumber \\
&=& 116\, 591\, 596(67)\times 10^{-11}.
\ee
Currently, the combined CERN \cite{PDG98} and BNL \cite{VernonBulg} 
measurements give
\be
a_\mu^{\rm exp} = 116\,592\,350(730) \times 10^{-11}.
\ee
It is expected that the ongoing run at BNL will reduce the uncertainty
by a factor of three and in the long term $\pm 40\times 10^{-11}$ is
achievable.  

Comparing experiment and theory, one finds
\be
 a_\mu^{\rm exp} - a_\mu^{\rm SM} \simeq (750 \pm 733) \times 10^{-11}
\ee
which implies good consistency; but leaves open the possibility of
``New Physics'' contributions to $a_\mu$ as large as $\sim {\cal
O}(10^{-8})$.  

We subsequently consider several examples of ``New Physics'' effects
on $a_\mu$.  To estimate the sensitivity of ongoing measurements and
theory, we assume $\pm 100\times 10^{-11}$ combined precision is
attainable.  However, if the theoretical error can be further reduced
and the experiment proceeds as planned, a reduction in that
uncertainty by 2--3 appears feasible.

\subsection{``New Physics'' and Electroweak Radiative Corrections}
``New Physics,'' beyond Standard Model expectations, will in general
give rise to additional $a_\mu$ contributions which we collectively
call $a_\mu^{\rm New \; Physics}$.  Before discussing specific examples,
we consider the electroweak radiative corrections to such ``New Physics''
effects.

Most ``New Physics'' effects contribute directly to the dimension 5
magnetic dipole operator.  In that case, they are subject to the same
EW suppression factor as the $W$ loop contribution to $a_\mu^{\rm
EW}$.  From the calculation in Ref.~\cite{CKM96}, one finds a leading
log suppression factor
\be
1-{4\alpha \over \pi} \ln {M\over m_\mu} 
\label{eq:sup}
\ee
where $M$ is the characteristic ``New Physics'' scale.  For
$M\sim 100$ GeV, that factor corresponds to about a 6.4\% reduction. 

\subsection{Supersymmetry}
The supersymmetric contributions to $a_\mu$ stem from
smuon--neutralino and sneutrino-chargino loops
\cite{Lopez:1994vi,Nath95,Moroi:1996yh,Carena:1997qa} (see Fig.~1). 
\begin{figure}[thb]
\hspace*{-2mm}
\begin{minipage}{16.cm}
\vspace*{12mm}
\[
\hspace*{0mm}
\mbox{ 
\begin{tabular}{cc}
\psfig{figure=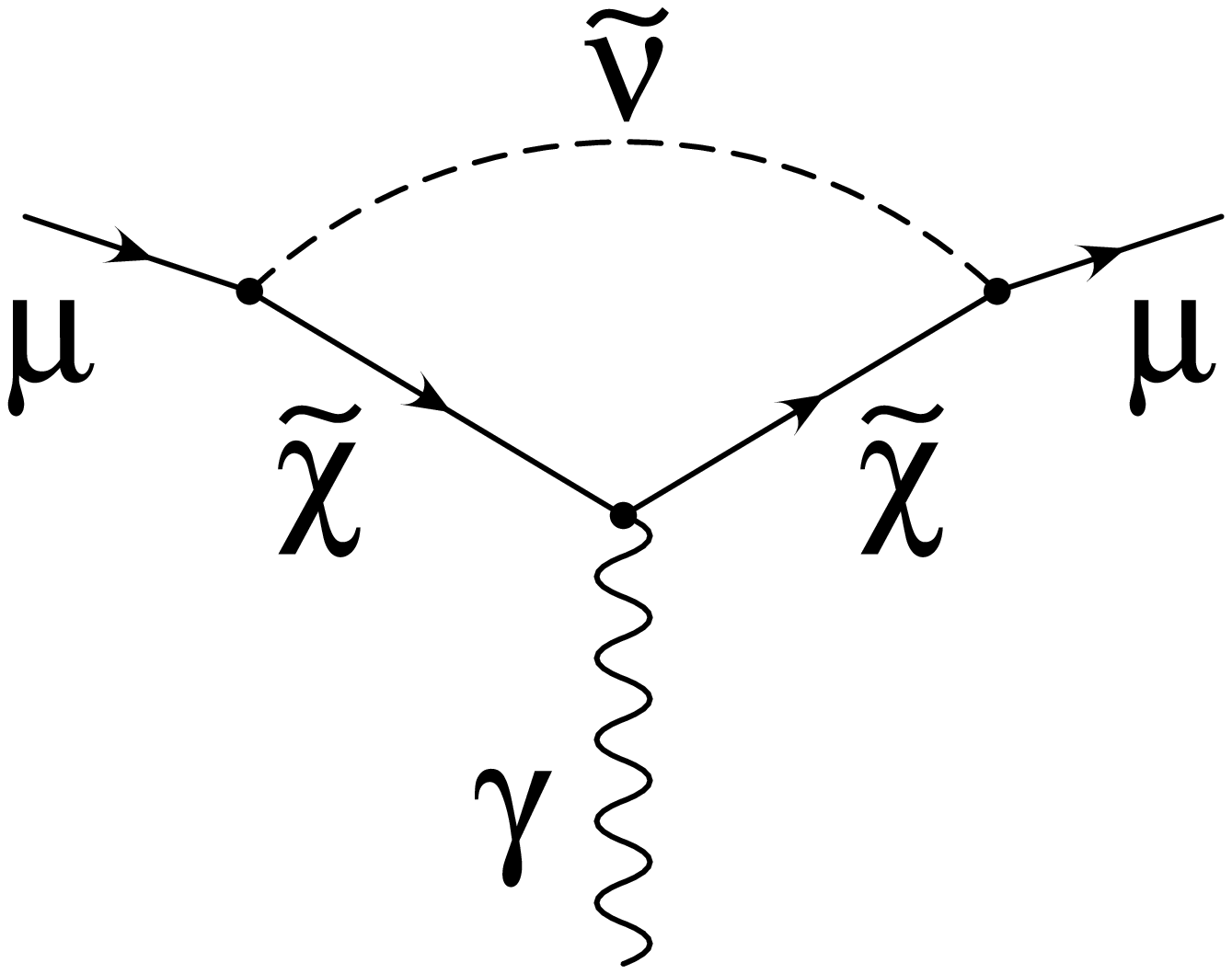,width=35mm,bbllx=72pt,bblly=291pt,%
bburx=544pt,bbury=540pt} 
& \hspace*{-2mm}
\psfig{figure=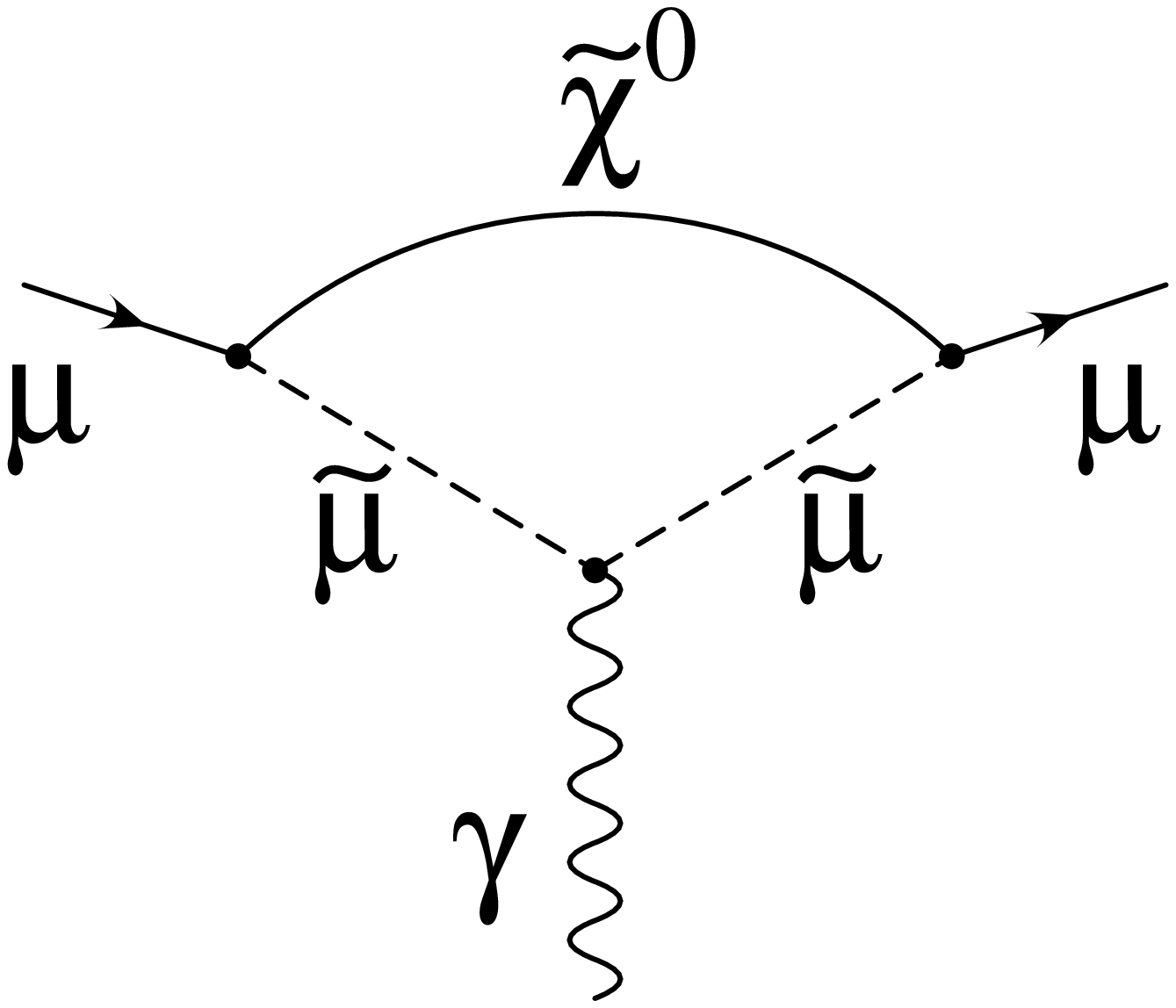,width=35mm,bbllx=72pt,bblly=291pt,%
bburx=544pt,bbury=540pt} 
\end{tabular}
}
\]
\end{minipage}
\caption{Supersymmetric loop contributing to the muon anomalous
magnetic moment.}
\label{fig1}
\end{figure}
They can be significant if the supersymmetric particles are not too
massive and if $\tan\beta\equiv v_2/v_1$ is large. Indeed, the one
loop effect is given in the large $\tan\beta$ limit by
\be
a_\mu^{\rm SUSY} \simeq {\alpha\over 8\pi\sin^2\theta_W}
{m_\mu^2\over \widetilde{m}^2}\tan\beta
\ee
where $\widetilde{m}$ represents a typical SUSY loop mass.  Including
the EW suppression factor in Eq.~(\ref{eq:sup}) then implies
\be
a_\mu^{\rm SUSY} \simeq 140\times 10^{-11} \left( 100 {\rm\ GeV}\over 
\widetilde{m}\right)^2 \tan\beta.
\ee
For large $\tan\beta\simeq 40$ scenarios, even an experimental
sensitivity of $\pm 100\times 10^{-11}$ probes $\widetilde{m}$ at the
750 GeV level and becomes competitive with direct high energy collider
searches. 

\subsection{Origin of Muon Mass}
In models where the muon mass is generated by quantum loops, similar
loop effects will also give additional contributions to $a_\mu$
\cite{MassMech}. 
Under very general assumptions, the induced $\delta a_\mu$ is given by 
\be
\delta a_\mu = C{m_\mu^2\over \Lambda^2}, \qquad C\sim {\cal O}(1),
\ee
where $\Lambda$ is the scale of ``New Physics'' responsible for
generating $m_\mu$.  Examples of such mechanisms include: extended
technicolor, multi-Higgs models, compositeness, etc.  For $a_\mu^{\rm
New \; Physics}$ sensitivity of $\pm 100 \times 10^{-11}$ and $C\sim
{\cal O}(1)$, $\Lambda \gsim 3$ TeV is probed.

\subsection{Anomalous $W$ couplings}
We generalize the $\gamma WW $ coupling such that 
the $W$ boson magnetic dipole moment is given by 
\be
\mu_W = {e\over 2m_W} (1+\kappa+\lambda)
\ee
and electric quadrupole moment by
\be
Q_W = -{e\over 2m_W} (\kappa - \lambda)
\ee
where $\kappa = 1$ and $\lambda = 0$ in the Standard Model,
i.e.~$g_W=2$. For non-standard couplings, one obtains the additional
one loop contribution to $a_\mu$ given by
\cite{Mery:1990dx,Herzog:1984nx,Suzuki:1985yh,Grau:1985zh} 
\be
a_\mu(\kappa,\lambda) \simeq
{G_\mu m_\mu^2\over 4\sqrt{2} \pi^2}
 \left[ (\kappa-1) \ln{\Lambda^2\over m_W^2} - {1\over
3}\lambda\right],
\ee
where $\Lambda$ is the high momentum cutoff required to give a finite
result. It presumably corresponds to the onset of ``New Physics'' such
as the $W$ compositeness scale, or new strong dynamics. Electroweak
effects reduce that contribution by roughly the suppression in
Eq.~(\ref{eq:sup}). Probing 
$a_\mu$ at the $\pm 100\times 10^{-11}$ level provides a sensitivity
to $|\kappa-1|$ of about $\pm 0.1$ (for $\Lambda\sim 1$ TeV) (see also
\cite{Renard:1997rg}).
Generally, one might expect $\kappa\sim(m_W/\Lambda)^2$
in theories with underlying strong dynamics at scale $\Lambda$. 

The ongoing experimental effort to improve the accuracy of $a_\mu$
measurement will improve the constraints on the ``New Physics''
scenarios we mentioned above, as well as on other theories and
phenomena, such as a general two-Higgs doublet model
\cite{Krawczyk:1997sm}, leptoquarks \cite{Couture95}, or four fermion
contact interactions \cite{Gonzalez-Garcia:1998ay} (see also
\cite{Stud98}).

\section{Tau}
The anomalous magnetic moment of the tau, $a_\tau$, is predicted to be
\be
a_\tau = a_\tau^{\rm QED}+a_\tau^{\rm had}+a_\tau^{\rm EW}
\ee
where \cite{Samuel:1991su}
\be
a_\tau^{\rm QED} &=& 1.1732\times 10^{-3}, \\
a_\tau^{\rm had} &=& 3.2(4)\times 10^{-6}, \\
a_\tau^{\rm EW} &=& 4.7\times 10^{-7}. 
\ee
We have updated those contributions to incorporate $m_\tau =
1.77705(26)$ GeV, a theoretical estimate of $a_\tau^{\rm had}$ from
perturbative QCD \cite{Hamzeh:1996wy} which is averaged with the
result given quoted in \cite{Samuel:1991su}, and two loop EW effects
\cite{CKM96} which suppress $a_\tau^{\rm EW}$ by about 15\%.  In
total, one finds
\be
a_\tau = 1.1769(4)\times 10^{-3}.
\ee
Experiments are currently not sensitive enough to measure $a_\tau$.
They can, however, indirectly bound an anomalously large $a_\tau$ due
to ``New Physics.''  For example, $e^+e^-\to \tau^+\tau^-$
cross-section measurements imply \cite{Silverman:1983ft}
\be
|a_\tau| \lsim 0.02.
\ee
Also, at this meeting L.~Taylor \cite{Taylor98} reported on analysis
of $e^+e^-\to \tau^+\tau^-\gamma$ by L3 and OPAL which gave a similar
constraint (see Table \ref{tab1}).  If ``New Physics'' effects in
$a_\tau$ are of the form $m_\tau^2/\Lambda^2$, then one would need to
extend such constraints to $|a_\tau| \lsim 0.0003$ to probe the
interesting regime $\Lambda\gsim 100$ GeV.

\subsection*{Acknowledgments}

We thank Professor T.~Kinoshita for informing us on the preliminary
value of $A_4$ in Eq.~(\ref{eq:eqed}).  A.C.~thanks Savely Karshenboim for
comments on \cite{Karshenboim:1993rt}, and the organizers of the
$\tau'98$ meeting, especially Antonio Pich and Alberto Ruiz, for
invitation and partial support of his participation.


\begin{thebibliography}{10}

\bibitem{Dehmelt87}
R.~S. {van Dyck Jr.}, P.~B. Schwinberg, and H.~G. Dehmelt, Phys. Rev. Lett.
  {\bf 59},  26  (1987).

\bibitem{PDG98}
C.~C. {\em et al.}~{(Particle Data Group)}, Eur. Phys. J. {\bf C3},  1  (1998).

\bibitem{VernonBulg}
V.~W. Hughes, to appear in Proc. of the Workshop on Frontier Tests of QED and
  Physics of the Vacuum, Sandansky, Bulgaria, June 1998.

\bibitem{Taylor98}
L. Taylor, hep-ph/9810463 (in these Proceedings).

\bibitem{Schwinger48}
J. Schwinger, Phys. Rev. {\bf 73},  416  (1948).

\bibitem{som57}
C.~M. Sommerfield, Phys. Rev. {\bf 107},  328  (1957);
 Ann. Phys. {\bf 5},  26  (1958).

\bibitem{pet57a}
A. Petermann, Nucl. Phys. {\bf 3},  689  (1957);
 Helv. Phys. Acta {\bf 30},  407  (1957).

\bibitem{Laporta:1996mq}
S. Laporta and E. Remiddi, Phys. Lett. {\bf B379},  283  (1996).

\bibitem{Kinoshita98}
T. Kinoshita, private communication.

\bibitem{Bar75}
R. Barbieri and E. Remiddi, Nucl. Phys. {\bf B90},  233  (1975).

\bibitem{Laporta:1993ju}
S. Laporta, Nuovo Cim. {\bf 106A},  675  (1993).

\bibitem{Laporta:1993pa}
S. Laporta and E. Remiddi, Phys. Lett. {\bf B301},  440  (1993).

\bibitem{Davier:1998si}
M. Davier and A. Hocker, hep-ph/9805470.

\bibitem{Krause:1997rf}
B. Krause, Phys. Lett. {\bf B390},  392  (1997).

\bibitem{CKM96}
A. Czarnecki, B. Krause, and W. Marciano, Phys. Rev. Lett. {\bf 76},  3267
  (1996).

\bibitem{Kinoshita:1996vz}
T. Kinoshita, Rept. Prog. Phys. {\bf 59},  1459  (1996).

\bibitem{Jeffery}
A. Jeffery {\it et~al.}, cited in \cite{Kinoshita:1996vz}.

\bibitem{Gab94}
G. Gabrielse and J. Tan,  in {\em Cavity Quantum Electrodynamics}, edited by
  P.~R. Berman (Academic Press, San Diego, 1994), p.\ 267.


\bibitem{Li:1993xf}
G. Li, R. Mendel, and M.~A. Samuel, Phys. Rev. {\bf D47},  1723  (1993).

\bibitem{Elend66}
H.~H. Elend, Phys. Lett. {\bf 20},  682  (1966), erratum: ibid., vol. 21, p.
  720.

\bibitem{Liu98}
W. Liu {\it et~al.}, submitted to Phys. Rev. Lett. 

\bibitem{Kinoshita:1993pq}
T. Kinoshita, Phys. Rev. {\bf D47},  5013  (1993).

\bibitem{Kinoshita:1990wp}
T. Kinoshita, B. Nizic, and Y. Okamoto, Phys. Rev. {\bf D41},  593  (1990).

\bibitem{Elkhovskii:1989cn}
A.~S. Yelkhovsky, Sov. J. Nucl. Phys. {\bf 49},  656  (1989).

\bibitem{Mil89}
A.~I. Milstein and A.~S. Yelkhovsky, Phys. Lett. {\bf B233},  11  (1989).

\bibitem{Karshenboim:1993rt}
S.~G. Karshenboim, Phys. Atom. Nucl. {\bf 56},  857  (1993).

\bibitem{Kataev:1992cp}
A.~L. Kataev, Phys. Lett. {\bf B284},  401  (1992).

\bibitem{Laporta:1994md}
S. Laporta, Phys. Lett. {\bf B328},  522  (1994).

\bibitem{Kataev:1995rw}
A.~L. Kataev and V.~V. Starshenko, Phys. Rev. {\bf D52},  402  (1995).

\bibitem{Ellis:1994qf}
J. Ellis, M. Karliner, M.~A. Samuel, and E. Steinfelds, hep-ph/9409376.


\bibitem{Broadhurst:1993si}
D.~J. Broadhurst, Z. Phys. {\bf C58},  339  (1993).

\bibitem{GRaf}
M. Gourdin and E. {de Rafael}, Nucl. Phys. {\bf B10},  667  (1669).

\bibitem{Jeg95}
S. Eidelman and F. Jegerlehner, Z. Phys. {\bf C67},  585  (1995).

\bibitem{Simon}
S. Eidelman, private communication.

\bibitem{Alemany:1997tn}
R. Alemany, M. Davier, and A. Hocker, Eur. Phys. J. {\bf C2},  123  (1998).

\bibitem{light}
E. de~Rafael, Phys. Lett. {\bf B322},  239  (1994).

\bibitem{Hayakawa:1998rq}
M. Hayakawa and T. Kinoshita, Phys. Rev. {\bf D57},  465  (1998).

\bibitem{Bijnens:1996xf}
J. Bijnens, E. Pallante, and J. Prades, Nucl. Phys. {\bf B474},  379  (1996).

\bibitem{OhtaPC}
S. Ohta, private communication.

\bibitem{fls72}
K. Fujikawa, B.~W. Lee, and A.~I. Sanda, Phys. Rev. {\bf D6},  2923  (1972).

\bibitem{Jackiw72}
R. Jackiw and S. Weinberg, Phys. Rev. {\bf D5},  2473  (1972).

\bibitem{ACM72}
G. Altarelli, N. Cabibbo, and L. Maiani, Phys. Lett. {\bf B40},  415  (1972).

\bibitem{Bars72}
I. Bars and M. Yoshimura, Phys. Rev. {\bf D6},  374  (1972).

\bibitem{Bardeen72}
W.~A. Bardeen, R. Gastmans, and B.~E. Lautrup, Nucl. Phys. {\bf B46},  315
  (1972).

\bibitem{CKM95}
A. Czarnecki, B. Krause, and W. Marciano, Phys. Rev. {\bf D52},  R2619  (1995).

\bibitem{KKSS}
T.~V. Kukhto, E.~A. Kuraev, A. Schiller, and Z.~K. Silagadze, Nucl. Phys. {\bf
  B371},  567  (1992).

\bibitem{Peris:1995bb}
S. Peris, M. Perrottet, and E. de~Rafael, Phys. Lett. {\bf B355},  523  (1995).

\bibitem{Degrassi:1998es}
G. Degrassi and G.~F. Giudice, Phys. Rev. {\bf D58},  053007  (1998).

\bibitem{Lopez:1994vi}
J.~L. Lopez, D.~V. Nanopoulos, and X. Wang, Phys. Rev. {\bf D49},  366  (1994).

\bibitem{Nath95}
U. Chattopadhyay and P. Nath, Phys. Rev. {\bf D53},  1648  (1996).

\bibitem{Moroi:1996yh}
T. Moroi, Phys. Rev. {\bf D53},  6565  (1996).

\bibitem{Carena:1997qa}
M. Carena, G.~F. Giudice, and C.~E.~M. Wagner, Phys. Lett. {\bf B390},  234
  (1997).

\bibitem{MassMech}
W. Marciano,  in {\em Particle Theory and Phenomenology}, edited by K. Lassila
  {\it et~al.} (World Scientific, Singapore, 1996), p.\ 22.

\bibitem{Mery:1990dx}
P. Mery, S.~E. Moubarik, M. Perrottet, and F.~M. Renard, Z. Phys. {\bf C46},
  229  (1990).

\bibitem{Herzog:1984nx}
F. Herzog, Phys. Lett. {\bf 148B},  355  (1984).

\bibitem{Suzuki:1985yh}
M. Suzuki, Phys. Lett. {\bf 153B},  289  (1985).

\bibitem{Grau:1985zh}
A. Grau and J.~A. Grifols, Phys. Lett. {\bf 154B},  283  (1985).

\bibitem{Renard:1997rg}
F.~M. Renard, S. Spagnolo, and C. Verzegnassi, Phys. Lett. {\bf B409},  398
  (1997).

\bibitem{Krawczyk:1997sm}
M. Krawczyk and J. Zochowski, Phys. Rev. {\bf D55},  6968  (1997).

\bibitem{Couture95}
G. Couture and H. K{\"o}nig, Phys. Rev. {\bf D53},  555  (1996).

\bibitem{Gonzalez-Garcia:1998ay}
M.~C. Gonzalez-Garcia, A. Gusso, and S.~F. Novaes, hep-ph/9802254.

\bibitem{Stud98}
A.~I. Studenikin, hep-ph/9808219.

\bibitem{Samuel:1991su}
M.~A. Samuel, G. Li, and R. Mendel, Phys. Rev. Lett. {\bf 67},  668  (1991),
  erratum: ibid. {\bf 69}, 995 (1992). See also S. Narison,
  J. Phys. {\bf G4}, 1849 (1978).

\bibitem{Hamzeh:1996wy}
F. Hamzeh and N.~F. Nasrallah, Phys. Lett. {\bf B373},  211  (1996).

\bibitem{Silverman:1983ft}
D.~J. Silverman and G.~L. Shaw, Phys. Rev. {\bf D27},  1196  (1983).

\end{thebibliography}

\end{document}